\documentclass[twoside,a4paper,11pt]{sea22}
\usepackage{graphicx}
\usepackage{hyperref}
\usepackage{blindtext}
\usepackage{media9}

\newcommand{\niiaur}{[N~{\sc ii}]~$\lambda$5755}
\newcommand{\nii}{N~{\sc ii}}
\newcommand{\oii}{O~{\sc ii}}
\newcommand{\hii}{H~{\sc ii}}
\newcommand{\sii}{S~{\sc ii}}
\newcommand{\siii}{S~{\sc iii}}
\newcommand{\cliii}{Cl~{\sc iii}}
\newcommand{\hi}{H~{\sc i}}

\begin{document}
\pagenumbering{arabic}
\pagestyle{myheadings}
\thispagestyle{empty}
{\flushleft\includegraphics[width=\textwidth,bb=58 650 590 680]{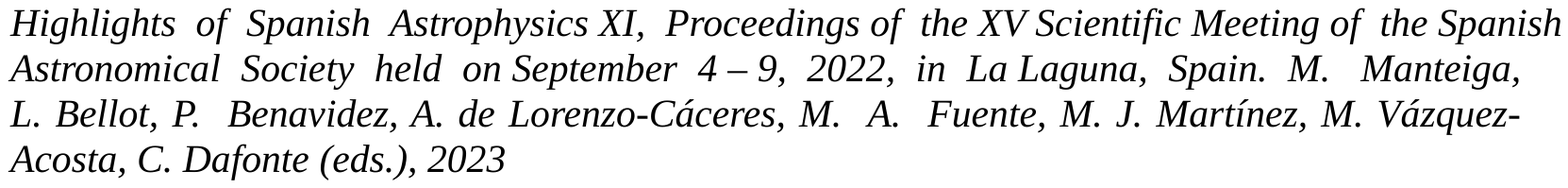}}
\vspace*{0.2cm}
\begin{flushleft}
{\bf {\LARGE
%
%%% TITLE of the paper. 
The planetary nebula NGC\,6153 through the eyes of MUSE
%
% Do not delete next few lines
}\\
\vspace*{1cm}
%
%%% Include here the LIST OF AUTHORS.
%%% Note that the last author has to be preceeded by an AND.
G\'omez-Llanos, V.$^{1,2}$,
Garc\'ia-Rojas, J.$^{1,2}$, 
Morisset, C.$^{3}$,
Jones, D.$^{1,2}$,
Monteiro, H.$^{4}$,
Wesson, R.$^{5}$,
Boffin, H. M. J.$^{6}$,
Corradi, R. L. M. $^{2,7}$,
P\'erez-Toledo, F.$^{7}$,
and
Rodr\'iguez-Gil, P.$^{1,2}$
%
% Do not delete next few lines
}\\
\vspace*{0.5cm}
%
%%% AFFILIATIONS LIST.
%%% and the AFFILIATIONS LIST. Note that one affiliation per line.
%%% Add as many affiliations as necessary. 
$^{1}$
Instituto de Astrof\'isica de Canarias, E-38205 La Laguna, Tenerife, Spain\\
$^{2}$
Departamento de Astrof\'isica, Universidad de La Laguna, E-38206 La Laguna, Tenerife, Spain\\
$^{3}$
Instituto de Astronom\'ia, UNAM, Apdo. postal 106, C.P. 22800 Ensenada, Baja California, M\'exico \\
$^{4}$
Instituto de F\'isica e Qu\'imica, Universidade Federal de Itajub\'a, Av. BPS 1303-Pinheirinho, 37500-903, Itajub\'a, Brazil \\
$^{5}$
School of Physics and Astronomy, Cardiff University, Queen's Buildings, The Parade, Cardiff CF24 3AA, UK \\
$^{6}$
European Southern Observatory, Karl-Schwarzschild-Str. 2, D-85738 Garching bei M\"unchen, Germany \\
$^{7}$
Gran Telescopio CANARIAS S.A., c/ Cuesta de San Jos\'e s/n, Bre\~na Baja, E-38712 Santa Cruz de Tenerife, Spain \\
%
% Do not delete next few lines
\end{flushleft}
%
% Headings
\markboth{
%%% Type the SHORT version of the paper title.
The planetary nebula NGC\,6153 through the eyes of MUSE
}{ % Do not delete
%
%%%  First Author \& Second Author   OR   First-author et al. if the author list 
%%% contains three or more authors.
G\'omez-Llanos et al. 
% 
% Do not delete next few lines
}
\thispagestyle{empty}
\vspace*{0.4cm}
\begin{minipage}[l]{0.09\textwidth}
\ 
\end{minipage}
\begin{minipage}[r]{0.9\textwidth}
\vspace{1cm}
\section*{Abstract}{\small
%
% ABSTRACT ABSTRACT ABSTRACT
%%% Type the ABSTRACT of your oral contribution or poster
In this contribution, we present the results of a study on the high abundance discrepancy factor (ADF $\sim$ 10) planetary nebula (PN) NGC\,6153 with MUSE. We have constructed flux maps for dozens of emission lines, that allowed us to build spatially resolved maps of extinction, electron temperature ($T_{\rm e}$), electron density ($n_{\rm e}$), and ionic abundances. We have simultaneously constructed ADF maps for O$^+$ and O$^{2+}$ and found that they centrally peak in this PN, with a remarkable spatial coincidence with the low $T_{\rm e}$ found from recombination line diagnostics. This finding strongly supports the hypothesis that two distinct gas phases co-exist: one cold and metal-rich, and a second warm and with ``normal'' metal content. We show that to build $T_{\rm e}$([N~{\sc ii}]) and ionic abundance maps of low-ionization species for these objects, recombination contribution to the auroral [N~{\sc ii}] and [O~{\sc ii}] lines must be properly evaluated and corrected.

%
% Do not delete next few lines
\normalsize}
\end{minipage}
%
%%% BODY of the paper
%
\section{Introduction \label{sec:intro}}

 Since \cite{1942Wyse} first reported it, the abundance discrepancy problem, i.~e., the long standing difference between the chemical abundances computed for a given metal ion from recombination lines (RLs) or collisionally excited lines (CELs) has cast doubt on the chemical abundances determinations in both planetary nebulae (PNe) and \hii\ regions. The RL/CEL ratio, the so-called abundance discrepancy factor (ADF) can show extreme values (up to 700) in some PNe. Some scenarios have been proposed to explain this problem (see \cite{2019Garcia-Rojas} and references therein). However, for PNe there are several observational evidences pointing to the presence of two gas components with different chemical composition (and possibly kinematics) (\cite{1982Barker, 2001Garnett, 2008Tsamis, 2013Richer, 2016Garcia-Rojas, 2017Pena, 2017Richer, 2022Richer}). This bi-abundance scenario (first proposed by \cite{1990Torres-Peimbert}) consists of a ``normal'' chemical composition gas with a relatively warm electron temperature ($T_{\rm e}\sim$10,000$\,$K) that emits mainly the metal CELs and the H and He RLs , and an H-poor gas with a much lower temperature ($\sim$1,000$\,$K) and higher density whose emission is dominated by metals RLs.

NGC~6153 is a southern PN with strong emission of metal RLs, thus making it a good object to address the abundance discrepancy problem (\cite{2000Liu, 2002Pequignot, 2008Tsamis, 2011Yuan, 2020Gomez-Llanos, 2022Richer}). The chemical composition of this PN has been extensively studied by different authors who hypothesised on the presence of two plasma components based on deep multi-wavelength spectroscopic data (\cite{2000Liu}) or on integral field spectroscopy (\cite{2008Tsamis}). Very recently,  making use of a very high-resolution spectra and position-velocity maps, \cite{2022Richer} reached a similar conclusion with the addition of finding also differences in the kinematics of the gas between the two components. From the theoretical side, empirical, one-dimensional, and 3D photoionization models have been constructed for this object considering a chemically inhomogeneous gas, successfully fitting both the RLs and CELs (\cite{2000Liu, 2002Pequignot, 2011Yuan, 2020Gomez-Llanos}) and hence, strengthening the bi-abundance scenario for this object.

The analysis of 2D spectroscopic data of PNe with large ADFs have revealed that extreme care should be taken when constructing physical conditions and ionic abundance maps, especially for low ionization species (see \cite{2022Garcia-Rojas}). In this work we present some of the preliminary results we have obtained from the analysis of very deep MUSE data of NGC\,6153.  

\section{Observations \label{sec:obs}}

NGC\,6153 was observed with the Multi Unit Spectroscopic Explorer (MUSE) integral-field spectrograph  (\cite{2010Bacon}) on the Very Large Telescope (VLT), in seeing-limited mode, on the night of 6 to 7 July 2016. We used the extended mode of MUSE (WFM-NOAO-E), which covers the wavelength range $460--930$\,nm with an effective spectral resolution that increases from $\mathcal{R} \sim 1600$ at the bluest wavelengths to $\mathcal{R} \sim 3500$ at the reddest wavelengths. The on-target exposure time was 2320 s divided in several long and short exposures. The observing conditions, observation techniques and reduction process have been described by \cite{2022Garcia-Rojas}. 

\section{Electron temperature maps \label{sec:temp}}

From the MUSE observations of NGC~6153, we have constructed flux maps and their uncertainties for more than 60 emission lines following the same methodology as described by \cite{2022Garcia-Rojas}. We then built spatially resolved maps of extinction, electron temperature ($T_{\rm e}$), electron density ($n_{\rm e}$), and ionic abundances. The $T_{\rm e}$ and $n_{\rm e}$ maps were obtained using different line ratios as diagnostics (e.g., [\nii]~$\lambda$5755/$\lambda$6548\footnote{Hereafter all wavelengths will be in \AA.} and [\siii]~$\lambda$6312/$\lambda$9069 for $T_{\rm e}$, and [\sii]~$\lambda$6731/$\lambda$6716 and [\cliii]~$\lambda$5538/$\lambda$5518 for $n_{\rm e}$). However, the diagnostics based on second-row elements such as O and N can have an important contribution from recombination to the low metastable levels, like \niiaur\ and [\oii]~$\lambda \lambda 7320,7730$, that if not corrected, will lead to an overestimate of the temperature. This may be especially important for spatially resolved observations of PNe with high ADF, where in extreme cases, the \niiaur\ emission can be dominated by recombination (see Fig.~7 in \cite{2022Garcia-Rojas}). 

\begin{figure}[]
    \centering
    \includegraphics[scale = 0.55]{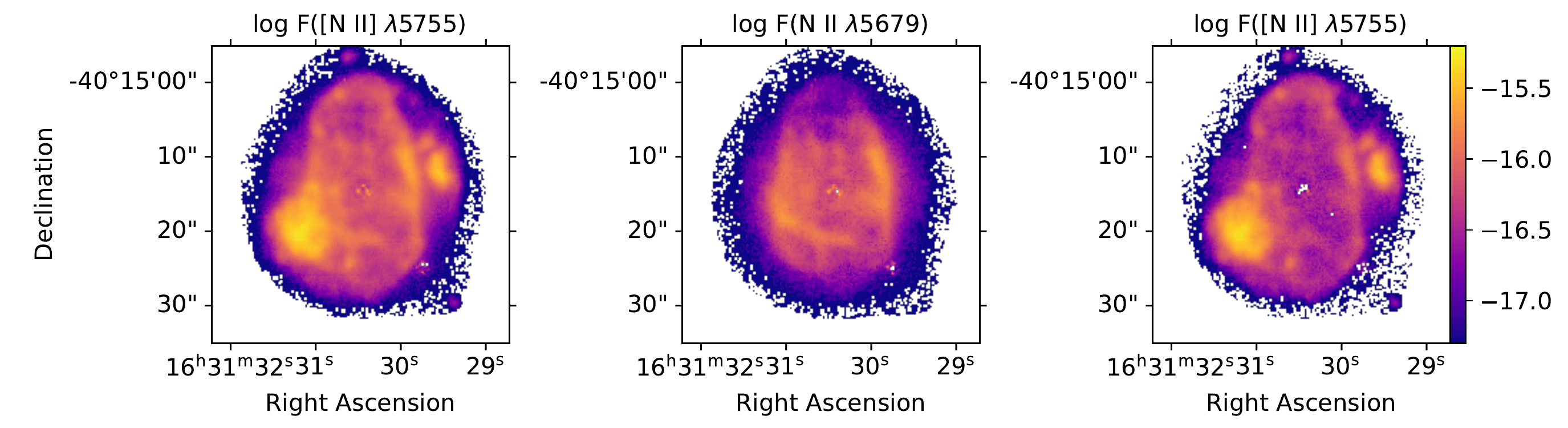}
    \caption{Left panel: spatial distribution of the auroral \niiaur\ emission line in the PN NGC\,6153 prior to applying the recombination contribution. Middle panel: spatial distribution of the \nii\ $\lambda$5679 RL. Right panel: same as left panel but after applying the recombination contribution correction, considering a constant $n_{\rm e} = 10^4\,\textrm{cm}^{-3}$ and $T_{\rm e} = 2,000\,\textrm{K}$ for the recombination emission.}
    \label{fig:N2_fluxes}
\end{figure}

\begin{figure}[]
    \centering
    \includegraphics[scale = 0.6]{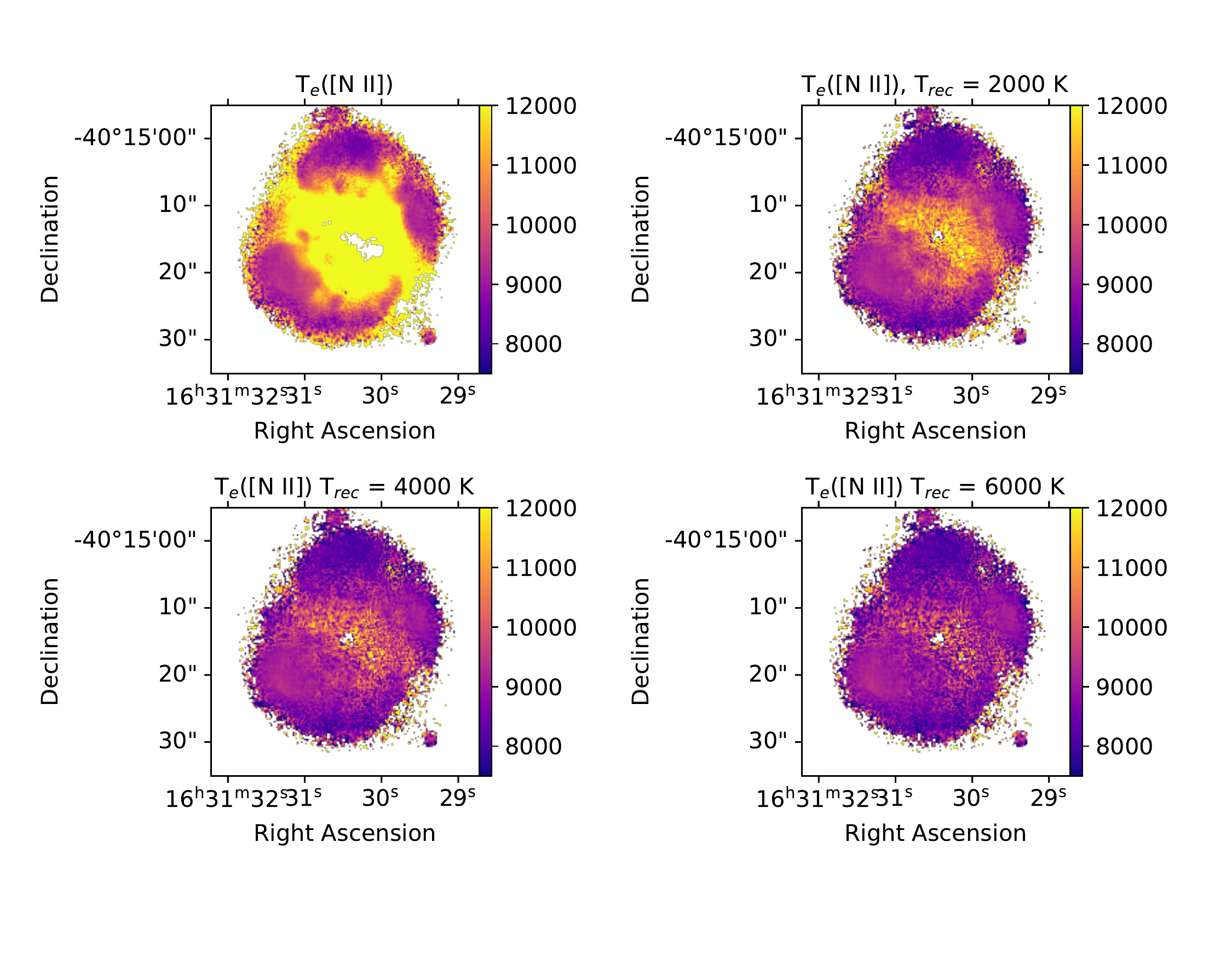}
    \caption{Electron temperature map computed from the $T_{\rm e}$-sensitive [\nii]~$\lambda \lambda$5755/6548 line ratio. In the top left panel we show the map with no recombination correction to the \niiaur\ auroral line; in the top right, bottom left, and bottom right panels we show the maps constructed with the recombination correction, considering a constant $n_{\rm e} = 10^4\,{\rm cm}^{-3}$ and $T_{\rm e} = 2,000\,\textrm{K}$, $4,000\,\textrm{K}$, and $6,000\,\textrm{K}$, respectively, for the recombination emission.}
    \label{fig:te_n2}
\end{figure}

In this work, we present the recombination contribution to \niiaur\ in the PN NGC~6153 (ADF$\sim$10 \cite{2000Liu, 2008Tsamis}). In the left panel of Fig.~\ref{fig:N2_fluxes} we show the spatially resolved emission of \niiaur. To estimate its recombination contribution we use Eq.~1 presented by \cite{2022Garcia-Rojas}, which is based on the emission of \nii~$\lambda$5679 (middle panel of Fig.~\ref{fig:N2_fluxes}) and the recombination emissivities of $j_{5755}(T_{\rm e}, n_{\rm e})$ and $j_{5679}(T_{\rm e}, n_{\rm e})$. \cite{2022Richer} presents an estimate of the electron temperature and density of the recombination emitting region in NGC~6153, giving an average value of $n_{\rm e} = 10^4\,\textrm{cm}^{-3}$ and $T_{\rm e} = 2,000\,\textrm{K}$. We use these values for the recombination emissivities. The corrected \niiaur\ emission map is presented in the right panel of Fig.~\ref{fig:N2_fluxes}, which shows the main emission in two bright knots and on the edges of the nebula's main shell, while the uncorrected flux also shows a bright emission at inner regions of the nebula. To emphasize the importance of this correction, in Fig.~\ref{fig:te_n2} we show the [\nii] $\lambda$5755/$\lambda$6548 temperature distribution map with and without applying the correction on the top right and left panels of Fig.~\ref{fig:te_n2}, respectively. Notice the considerably higher temperatures in the inner parts of the nebula that are predicted without the correction. We also explore the effect of increasing the temperature of the recombination emitting region to 4,000~K and 6,000~K (bottom left and bottom right panels of Fig.~\ref{fig:te_n2}), which results in a decrease on the temperature in the inner parts of the nebula.  

We tried to compute $T_{\rm e}$ and $n_{\rm e}$ from metal RL diagnostics in order to break the degeneracies found by \cite{2022Garcia-Rojas} when trying to fully characterize the H-poor component. However, the most sensitive {\oii} and {\nii} RLs are either out of the wavelength range covered by MUSE or the maps constructed were too noisy to reach any conclusion. As already pointed out, \cite{2022Richer} could estimate the physical conditions of the H-poor component of the gas where most of the metals recombination emission comes. These values can be used to qualitatively estimate the influence of the cold region on abundance determinations and to determine the oxygen mass ratio between the cold and the warm regions (G\'omez-Llanos et al. in prep.)   

\section{The abundance discrepancy maps \label{sec:adf_map}}

Once we computed the ionic abundances from CELs and RLs, we constructed the abundance discrepancy factor maps for O$^+$ and O$^{2+}$ following the methodology described by \cite{2022Garcia-Rojas}. \cite{2008Tsamis} constructed the ADF(O$^{2+}$) map for NGC\,6153, but only sampled a small area of the nebula. \cite{2022Richer} made a study on the variation of the ADF over a position-velocity map, finding the highest values at positions close to the central star. These authors also find that the ADF was close to the unity in the diffuse emission beyond the receding side of the main shell of the nebula. As far as we know, this is the first time that the ADF is mapped for the whole nebula.  

On the other hand, we have constructed the {\hi} RL temperature diagnostic from the ratio of the Paschen jump to the {\hi} P9 $\lambda$9229 line, following the same methodology described by \cite{2022Garcia-Rojas}. This temperature diagnostic can provide hints on the influence of the cold gas on the computation of the global physical conditions in the nebula (G\'omez-Llanos et al. in prep.). 

\begin{figure}[]
    \centering
    \includegraphics[scale = 0.4]{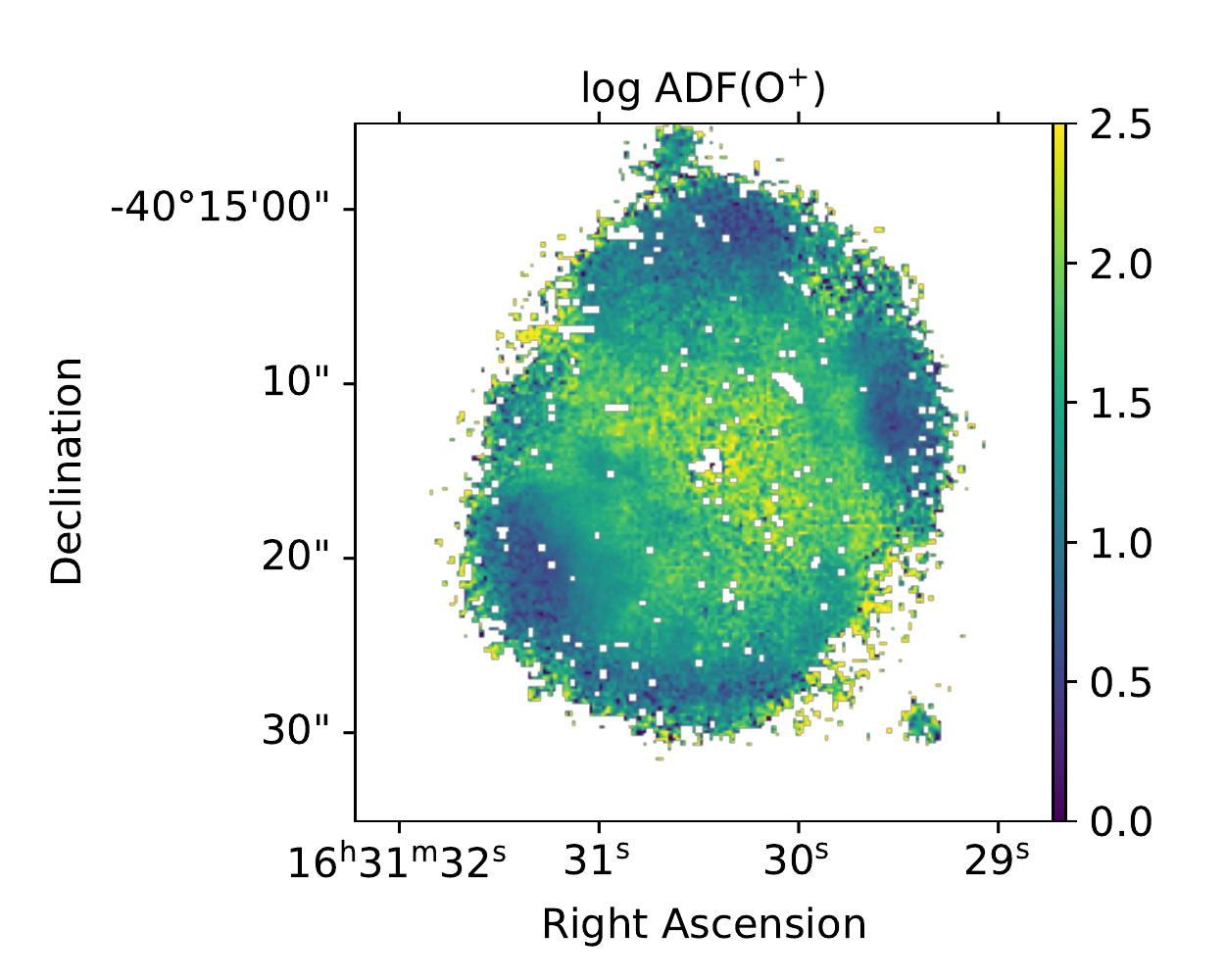}
    \includegraphics[scale = 0.4]{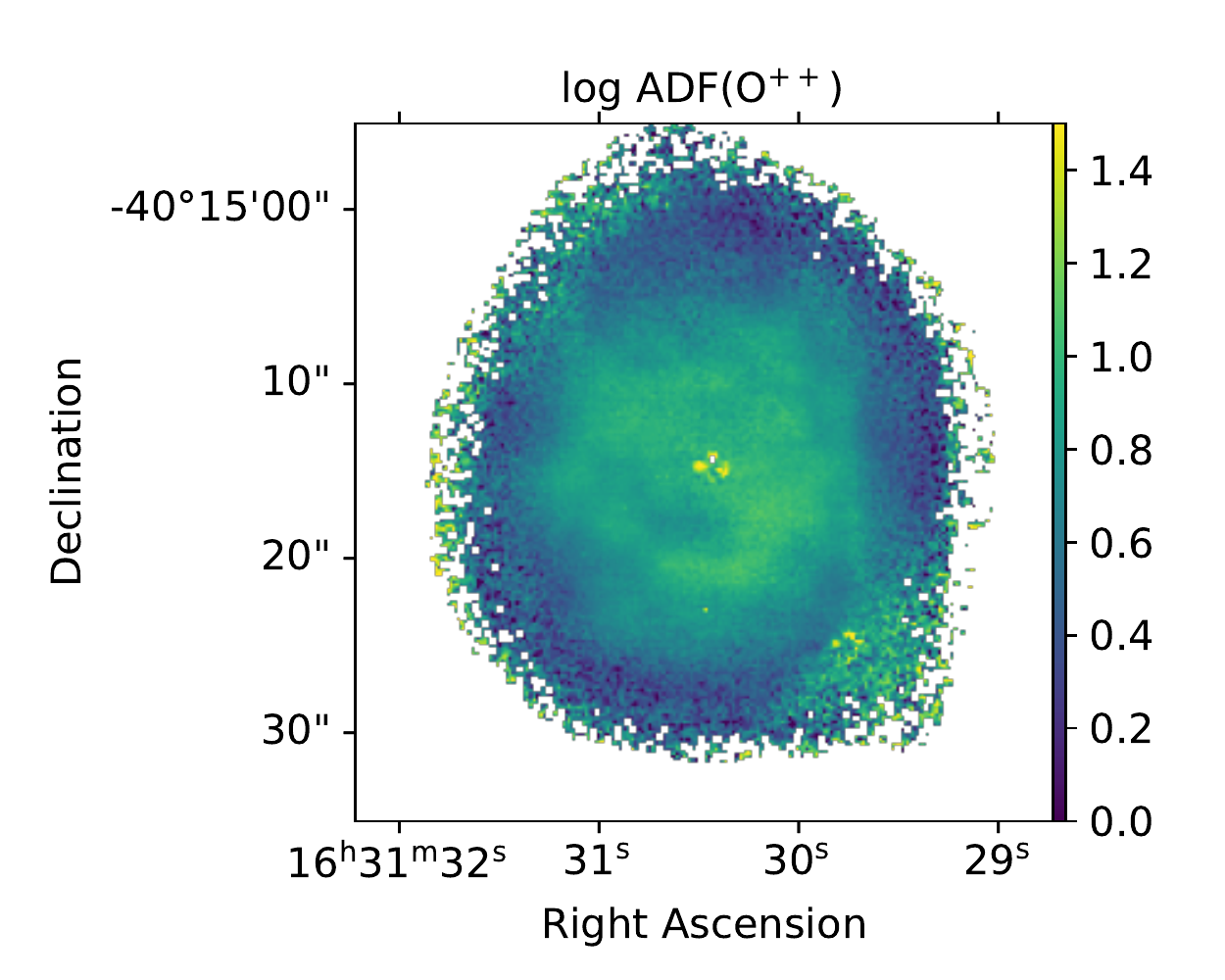}
    \includegraphics[scale = 0.4]{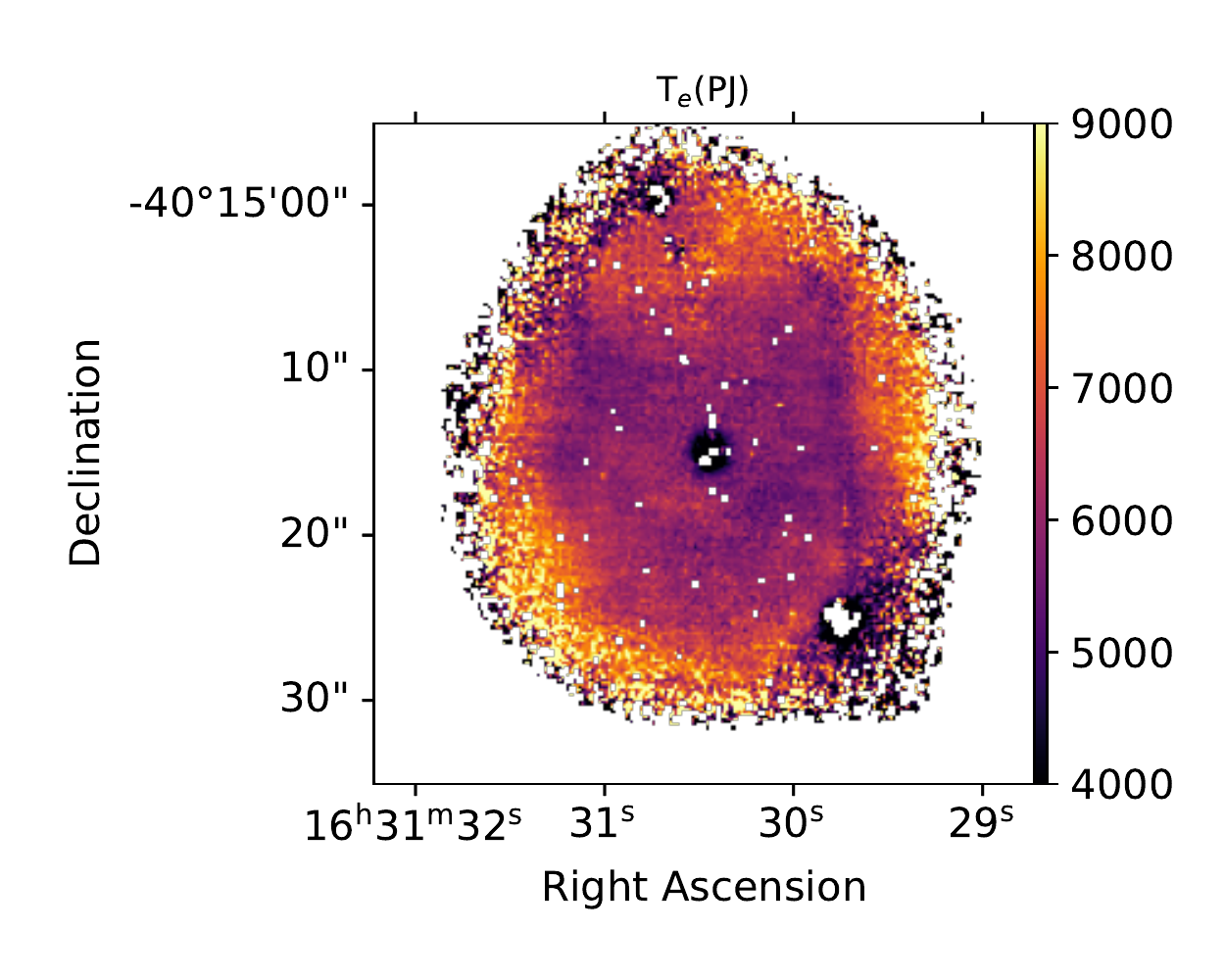}
    \caption{Spatial distributions of log[ADF(O$^+$)] (left panel) and log[ADF(O$^{2+}$)] (central panel). In the right panel we show the $T_{\rm e}$ map obtained from the Paschen jump relative to {\hi} P9 $\lambda$9229 line.} 
    \label{fig:adf}
\end{figure}

In Fig.~\ref{fig:adf} we present the spatial distribution of  log[ADF(O$^+$)] (left panel) and log[ADF(O$^{2+}$)] (central panel). In both maps, the ADF peak is clear in the central parts of the nebula, although the ADF variation in these central zones is not as extreme as that in the three high-ADF PNe presented in \cite{2022Garcia-Rojas}. This is an expected behaviour if we take into account the lower ADF value in the integrated spectrum of NGC\,6153. In the right panel of Fig.~\ref{fig:adf}, we illustrate the spatial distribution of the $T_{\rm e}$ obtained from the {\hi} Paschen jump. The spatial coincidence between the high values of the ADF(O$^{2+}$) and the low $T_{\rm e}$'s in this map is remarkable. This behavior strongly supports the hypothesis of the presence of a cold, metal-rich gas phase embedded in a warm gas phase with``normal'' metal content. The full analysis of the NGC\,6153 MUSE data set will be presented in a forthcoming paper (G\'omez-Llanos et al., in prep.).

%
% Do not delete the next line
\small  % Do not delete
%
%%% Comment the following line if you do not have acknowledgments.
\section*{Acknowledgements}   % Do not delete if you declare acknowledgments
This paper is based on observations made with ESO Telescopes at the Paranal Observatory under program ID 097.D-0241. VG-LL and JG-R acknowledge financial support from the Canarian Agency for Research, Innovation and Information Society (ACIISI), of the Canary Islands Government, and the European Regional Development Fund (ERDF), under grant with reference ProID2021010074. JG-R acknowledges support from an Advanced Fellowship under the Severo Ochoa excellence program CEX2019-000920-S. DJ acknowledges support from the Erasmus+ program of the European Union under grant number 2020–1–CZ01–KA203–078200. VG-LL, JG-R, DJ, and RC acknowledge support under grant P/308614 financed by funds transferred from the Spanish Ministerio de Ciencia, Innovaci\'on y Universidades, charged to the General State Budgets and with funds transferred from the General Budgets of the Autonomous Community of the Canary Islands by the MCIU. CM acknowledges support from grant UNAM / PAPIIT –IN101220. PR-G acknowledges support from the Consejer\'ia de Econom\'ia, Conocimiento y Empleo del Gobierno de Canarias and the European Regional Development Fund (ERDF) under grant with reference ProID2021010132 and ProID2020010104.
%
% Do not delete the next few lines

%
\end{document}